\documentclass{ws-ijmpc}

\begin{document}

\markboth{Carmen Pellicer-Lostao, Ricardo Lopez-Ruiz}
{Instructions for Typing Manuscripts (Paper's Title)}


\title{TRANSITION FROM EXPONENTIAL TO POWER LAW INCOME DISTRIBUTIONS IN A CHAOTIC MARKET}

\author{CARMEN PELLICER-LOSTAO}
\address{Department of Computer Science and BIFI, University of Zaragoza\\Centro Politecnico 
Superior, Campus R\'{\i}o Ebro - Maria de Luna 3,\\ZARAGOZA (50015), SPAIN.\\carmen.pellicer@unizar.es}

\author{RICARDO LOPEZ-RUIZ}
\address{Department of Computer Science and BIFI, University of Zaragoza\\Faculty of Science, 
Campus San Francisco, Building B,\\
ZARAGOZA (50009), SPAIN.\\rilopez@unizar.es}



\maketitle

\begin{history}
\end{history}

\begin{abstract}
Economy is demanding new models, able to understand and predict the evolution of markets. 
To this respect, Econophysics offers models of markets as complex systems, that try to 
comprehend macro-, system-wide states of the economy from the interaction of many agents 
at micro-level. One of these models is the gas-like model for trading markets. This tries 
to predict money distributions in closed economies and quite simply, obtains the ones 
observed in real economies. However, it reveals technical hitches to explain the power law 
distribution, observed in individuals with high incomes. In this work, non linear dynamics 
is introduced in the gas-like model in way that an effort to overcome these flaws. 
A particular chaotic dynamics is used to break the pairing symmetry of agents 
$(i,j)\Leftrightarrow(j,i)$. The results demonstrate that a "chaotic gas-like model" can 
reproduce the Exponential and Power law distributions observed in real economies. Moreover, 
it controls the transition between them. This may give some insight of the micro-level causes 
that originate unfair distributions of money in a global society. Ultimately, the chaotic 
model makes obvious the inherent instability of asymmetric scenarios, where sinks of wealth 
appear and doom the market to extreme inequality.

\end{abstract}

\ccode{PACS Nos.: 11.25.Hf, 123.1K}

\section{INTRODUCTION}
After the economic crisis of 2008, much criticism has been thrown upon modern economic theories. 
They have failed to predict the financial crisis and foresee its depthless [~\refcite{1}]. 
On the whole, this failure has been charged to the inherent complexity of the markets [~\refcite{2}].

In this respect, Econophysics has been ultimately offering new tools and perspectives to deal with 
economic complexity [~\refcite{3} -~\refcite{5}]. Simple stochastic models have been developed, 
where many agents interact at micro level giving rise to global empirical regularities observed 
in real markets. These models are giving some guidance to uncover the underlying rules of real economy.

One of the most relevant examples of these models is the conjecture of a kinetic theory with 
(ideal) gas-like behavior for trading markets [~\refcite{4} -~\refcite{9}]. This model offers 
a simple scheme to predict money distributions in a closed economic community of individuals. 
In it, each agent is identified as a gas molecule that interacts randomly with others, trading 
in elastic or money-conservative collisions. Randomness is also an essential ingredient of this 
model, for agents interact in pairs chosen at random and exchange a random quantity of money. 
In the end, the model shows that the asymptotic distribution of money in the community will 
follow the exponential (Boltzmann-Gibbs) law for a wide variety of trading rules [~\refcite{5}].

This result may elucidate real economic data, for nowadays it is well established that income 
and wealth distributions show one phase of exponential profile that covers about $90-95\%$ of 
individuals (low and medium incomes) [~\refcite{10} -~\refcite{12}]. Despite of this fact, the 
model shows some technical hitches to explain the other phase observed in real economies. This 
is the power law (Pareto) distribution profile integrated by the individuals with high incomes 
[~\refcite{12} -~\refcite{15}]. The model needs to introduce additional elements to obtain this 
distribution, such as saving propensity among traders or diffusion theory [~\refcite{5} ,~\refcite{9}].

The work presented here, intends to contribute in this respect. To do that, it proposes to 
incorporate chaotic dynamics in the traditional gas-like models. This is done upon two facts 
that seem particularly relevant in this purpose. The first one is that, determinism and 
unpredictability, the two essential components of chaotic systems, take part in the evolution 
of Economy and Financial Markets. On one hand, there is some evidence of markets being not purely 
random, for most economic transactions are driven by some specific interest (of profit) between 
the interacting parts. On the other hand, real economy shows periodically its unpredictable 
component with recurrent crisis. The prediction of the future situation of an economic system 
resembles somehow the weather prediction. Therefore, it can be sustained that market dynamics 
appear to be chaotic; in the short-time they evolve under deterministic forces though, in the 
long term, these kind of systems show an inherent instability.

The second fact is that the transition from the Boltzmann-Gibbs to the Pareto distribution may 
require the introduction of some kind of inhomogeneity that breaks the random indistinguishability 
among the individuals in the market. Though this is not a necessary condition, it may produce it. 
By now there have been different approaches that obtain this kind of transition, in [~\refcite{16}] 
a random market with inhomogeneous saving propensity obtains the Pareto distribution. The use of 
deterministic dynamics for modeling of financial markets has also been undertaken [~\refcite{17}] 
and deterministic dynamics with homogeneous models are also able to obtain the Pareto distribution 
[~\refcite{18}]. In this work, a deterministic model with asymmetric chaotic transactions is considered. 
The dynamics that governs the market is going to be selected so that is able of breaking the pairing 
symmetry of interacting agents $(i,j)\Leftrightarrow(j,i)$. A chaotic map with parametric-controlled 
symmetry may be an ideal candidate to obtain that, as it can quite simply produce parametric-controlled 
routines of evolution for the market. Additionally, a chaotic system will also represent a simple 
mechanism to introduce two other important ingredients: some degree of correlation between agents 
and establishing a complex pattern of transactions.

These concepts or hypothesis have inspired the introduction of a model for trading markets with 
chaotic patterns of evolution. Amazingly, it will be seen that a chaotic market is able to reproduce 
the two characteristic phases observed in real wealth distributions. The aim of this work is to 
observe what may happen at a micro level in the market, which is responsible of producing these 
two global phases.

This paper is organized as follows: section 2 introduces the basic theory of the gas-like model 
and describes the simulation scenario used in the computer simulations. Section 3 shows the results 
obtained in these simulations. Final section gathers the main conclusions and remarks obtained from this work.

\section{SIMULATION SCENARIO OF THE CHAOTIC MARKET}
The model proposed here is a multi-agent gas-like scenario [~\refcite{5}]. The study of these 
scenarios was first proposed by the authors in [~\refcite{19}]. There, it is shown that the use 
of chaotic numbers produces the exponential as well as other wealth distributions depending on 
how they are injected to the system. This paper considers the scenario where the selection of 
agents is chaotic, while the money exchanged at each interaction is a random quantity.

As in the gas-like model scenario a community of $N$ agents is given an initial equal quantity 
of money, $m_0$. The total amount of money, $M=N*m_0$, is conserved. The system evolves for a 
total time of $T=2*N^2$ transactions to reach the asymptotic equilibrium. For each transaction, 
at a given instant $t$, a pair of agents $(i,j)$ with money $({m_i}^t , {m_j}^t)$ is selected 
chaotically and a random amount of money $\Delta m$ is traded between them. The amount $\Delta m$ 
is obtained through equations (1) where $\upsilon$ is a float number in the interval $[0,1]$ 
produced by a standard random number generator:

\begin{eqnarray}
\begin{array}{lcl}
\Delta m=\upsilon*({m_i}^t+{m_j}^t)/2, \\[8pt]
{m_i}^{t+1}={m_i}^t-\Delta m, \\[8pt]
{m_j}^{t+1}={m_j}^t+\Delta m. \\[8pt]
\end{array}
\label{tradeRule}
\end{eqnarray}

This particular rule of trade is selected for simplicity and extensive use. Comparisons can be 
established with popularly referenced literature [~\refcite{4} -~\refcite{9}]. Here, the transaction 
of money is quite asymmetric as agent $j$ wins the money that $i$ losses. Also, if agent $i$ has 
not enough money $({m_i}^t < \Delta m)$, no transfer takes place.

\begin{figure}
  \centerline{
    \includegraphics[width=0.5\textwidth]{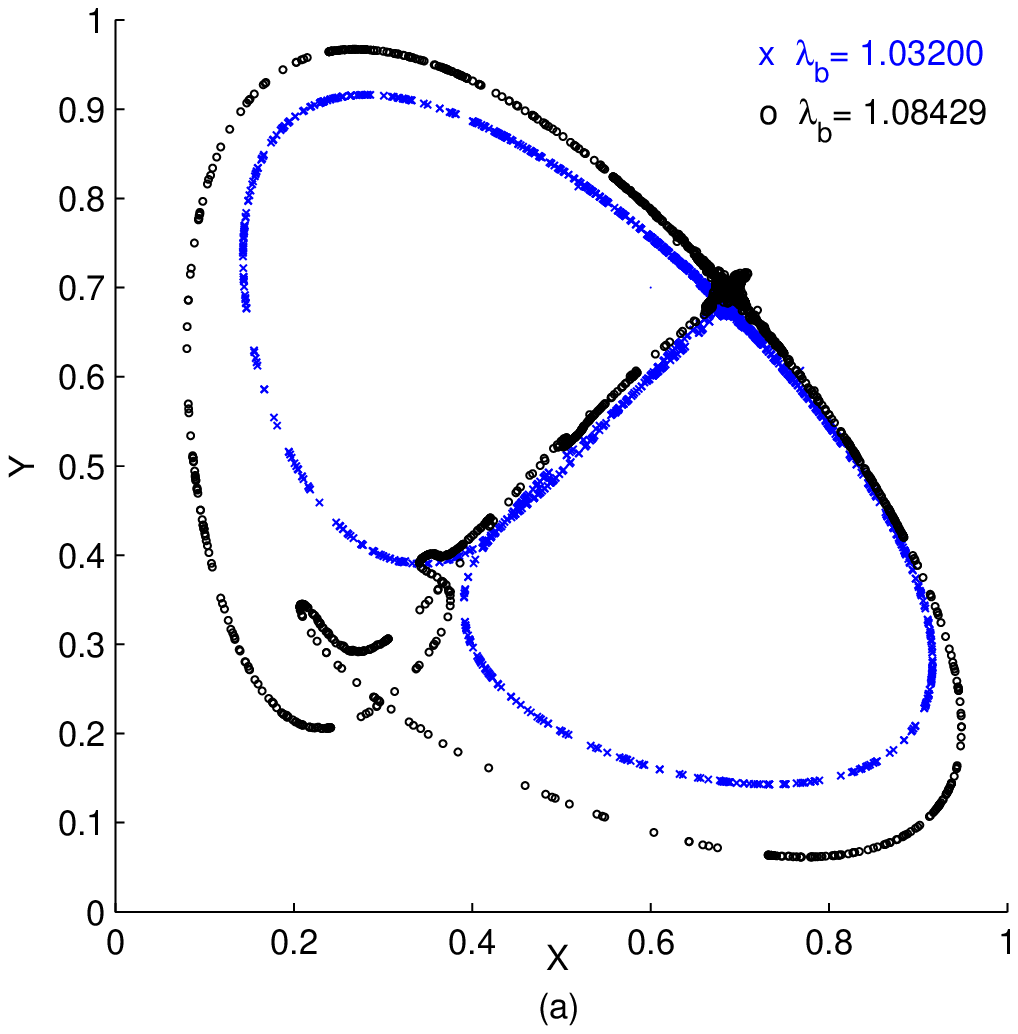} 
	\hskip 3mm \includegraphics[width=0.5\textwidth]{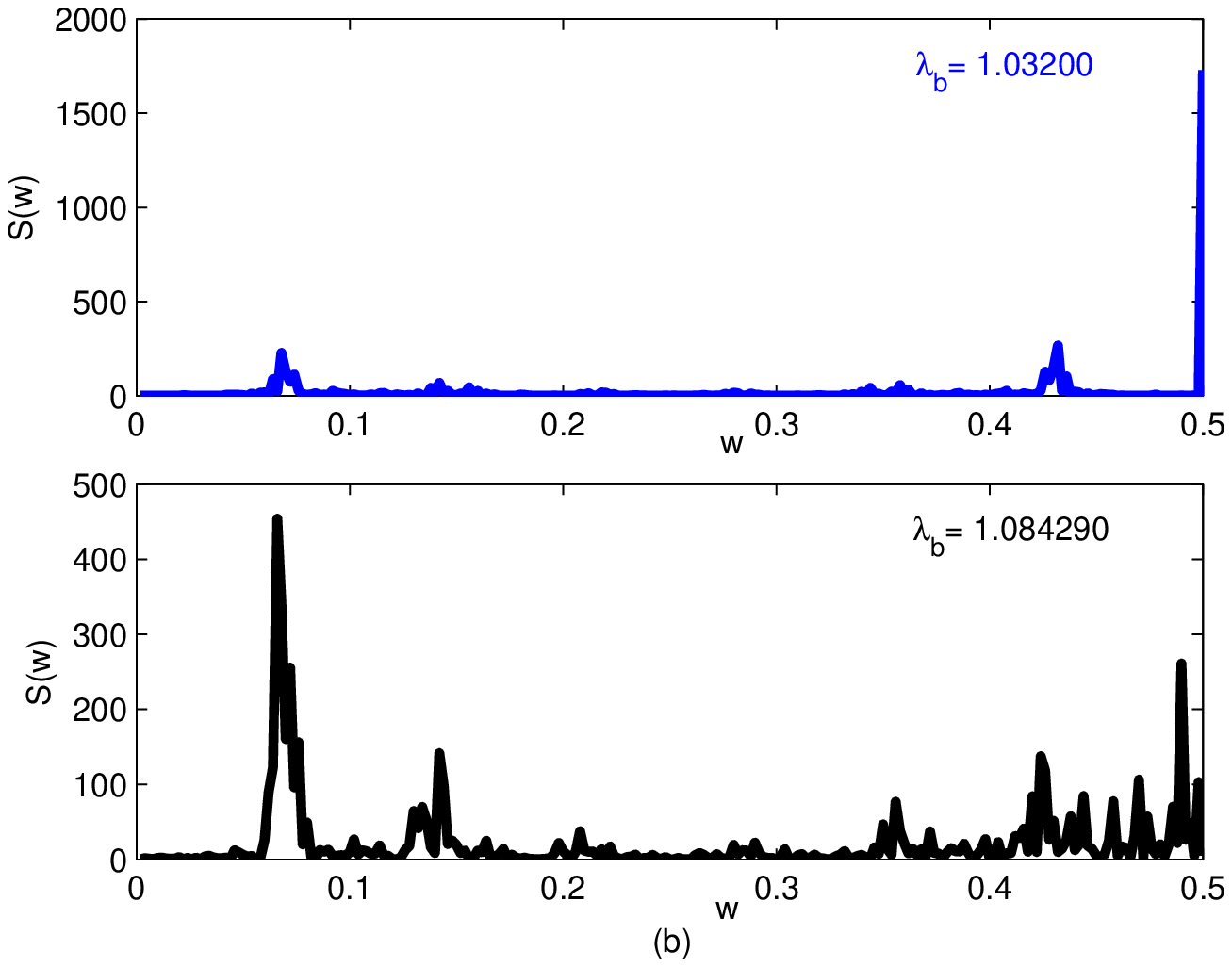}}
  \caption{(a) Representation of 2000 points of the chaotic attractor of the Logistic bimap. 
  This bimap is symmetric respective to the diagonal y=x  when $\lambda_a = \lambda_b$ . 
  (b) Representation of the spectrum of coordinate $x_t$ .The values of the parameters for 
  both graphics are $\lambda_a= \lambda_b= 1.032$ in the blue curves and $\lambda_a=1.032$, 
  $\lambda_b=1.084290$ in the black curves. The curves show the symmetric case, 
  $\lambda_a= \lambda_b= 1.032$ and the most asymmetric one, while chaotic behaviour is still 
  observed, $\lambda_a=1.032$, $\lambda_b=1.084290$}
  \label{Fig1}
\end{figure}

The chaotic selection of agents  $(i,j)$ for each interaction, is obtained by a particular 2D 
chaotic system, the Logistic bimap, described by L\'{o}pez-Ruiz and P\'{e}rez-Garc\'{\i}a in 
[~\refcite{20}]. This system is given by the following equations and depicted in Fig.\ref{Fig1}:

\begin{eqnarray}
\begin{array}{lcl}
T:[0,1]\times[0,1]\longrightarrow[0,1]\times[0,1] \\[8pt]
x_t =\lambda_a(3y_{t-1}+1)x_{t-1}(1-x_{t-1}), \\[8pt]
y_t=\lambda_b(3x_{t-1}+1)y_{t-1}(1-y_{t-1}).
\end{array}
\label{system}
\end{eqnarray}

The dynamical properties of this system make it an ideal candidate for our proposes. 
The model presented here requires a chaotic system able to break the pairing symmetry of 
interacting agents $(i,j)\Leftrightarrow(j,i)$. As it is explained below, the symmetry of 
this chaotic map can be controlled parametrically and so, this will allow to produce series 
of symmetric or asymmetric agents interactions at will.

The chaotic system of equation \ref{system} considers two Logistic maps that evolve in 
a coupled way. This system is used to select the interacting agents at a given time. 
The pair $(i,j)$ is easily obtained from the coordinates of a chaotic point at instant 
$t$, $X_t =[x_t, y_t]$, by a simple float to integer conversion:
\begin{eqnarray}
\begin{array}{lcl}
i=(int)(x_t*N), \\[8pt]
j=(int)(y_t*N).
\end{array}
\label{agents}
\end{eqnarray}

This procedure will make that the selection of a winner ($j$), or a looser ($i$), follows a 
chaotic pattern. Moreover, as there are two Logistic maps are coupled with each other, this 
bimap is able to introduce a strong correlation in the selection of agents of each group. 
This correlation is modulated or governed by the parameters $\lambda$. The Logistic bimap 
presents a chaotic attractor in the interval $\lambda_{a,b}\in[1.032, 1.0843 ]$ 
(see a real-time animation in [~\refcite{21}]). Consequently, the chaotic selection of agents 
is guaranteed by taking some appropriate $\lambda_{a,b}$ in this range. Fig.\ref{Fig1} 
(a) shows two trajectories for two different groups of values of $\lambda_a$ and $\lambda_b$, 
showing maximum asymmetry while prevailing chaotic behaviour.

A property of the Logistic bimap is its symmetry respective to the diagonal $y = x$ , 
when $\lambda_a = \lambda_b$ (see Fig.\ref{Fig1} (a)). The spectrum of coordinate $x_t$ 
shows a peak for $w=0.5$ (see Fig.\ref{Fig1} (b)) presenting an oscillation of period two 
that makes a jump over the diagonal axis alternatively between consecutive instants of time. 
Both sub-spaces $x>y$ and $y>x$ are visited with the same frequency and the shape of the attractor 
is symmetric. When $\lambda_b$ becomes greater than $\lambda_a$ the part of the attractor in 
sub-space $x<y$ becomes wider and the frequency of visits of each sub-space becomes different. 
These characteristics are used in the model as an input variable, so the degree of symmetry in 
the selection of agents can be modified at will.

In the end, the dynamics of the economic interactions are going to be complex and quite different 
from the random scenario, where any agent may interact with any other with the same probability. 
Here, market transactions are restricted in the sense that one agent will only interact with 
specific groups of other agents. To see this, just think that an $i^0$ agent given by an $x=x_0$ 
coordinate, with $x_0$ any constant value in the interval $[0,1]$. Draw the line $x=x_0$ 
in Fig.\ref{Fig1} (a) and then it will be seen that this agent will interact with its neighbors 
on the diagonal ($j~i$) and maybe with two other distant groups of agents. Additionally these 
secondary groups interact with other groups and so on, giving rise to a complex flow of interactions 
in the whole market. This may seem more realistic than the random case, as normally one individual 
doesn't interact directly with any other, but with specific groups of traders, being the complex 
connections of the community, the ones that develop a global indirect trading.

\section{SIMULATION RESULTS OBTAINED IN THE CHAOTIC GAS-LIKE MODEL}
Taking into account the chaotic market model discussed in the previous section, different computer 
simulations are carried out. In these simulations a community of $N=5000$ agents with initial money 
of $m_0=1000\$$  is taken. The simulations take a total time of $T=2*N^2 = 50$ millions of transactions.

Different cases are produced when different values of the chaotic parameters $\lambda_a$ and 
$\lambda_b$. In this way the symmetry of the selection of agents can be varied at will. This 
will allow us to see the effects of introducing an asymmetry in the process of selection of 
interacting agents. Simulations are produced when $\lambda_a =1.032$ has a fixed value and 
$\lambda_b$ varies from $1.032$ to $1.033162$ with increments of $\delta= 0.581{\times}10^{-4}$.

The results of these simulations reveal very interesting features. First, there is a group of 
individuals that keep their initial money and don't interact at all. This result would be 
consistent with real markets where not every agent are active. This number is due to the shape 
of the chaotic attractor. In Fig.\ref{Fig1} (a), it can be seen that the attractor hardly reaches 
the extreme values of the interval $[0,1]$. In the case of  $\lambda_a= \lambda_b = 1.032$ the 
number of non interacting agents is $1133$ in a community of $N=5000$. When $\lambda_b$ increases 
the attractor expands and this number becomes smaller.

\begin{figure}
  \centerline{
    \includegraphics[width=0.5\textwidth]{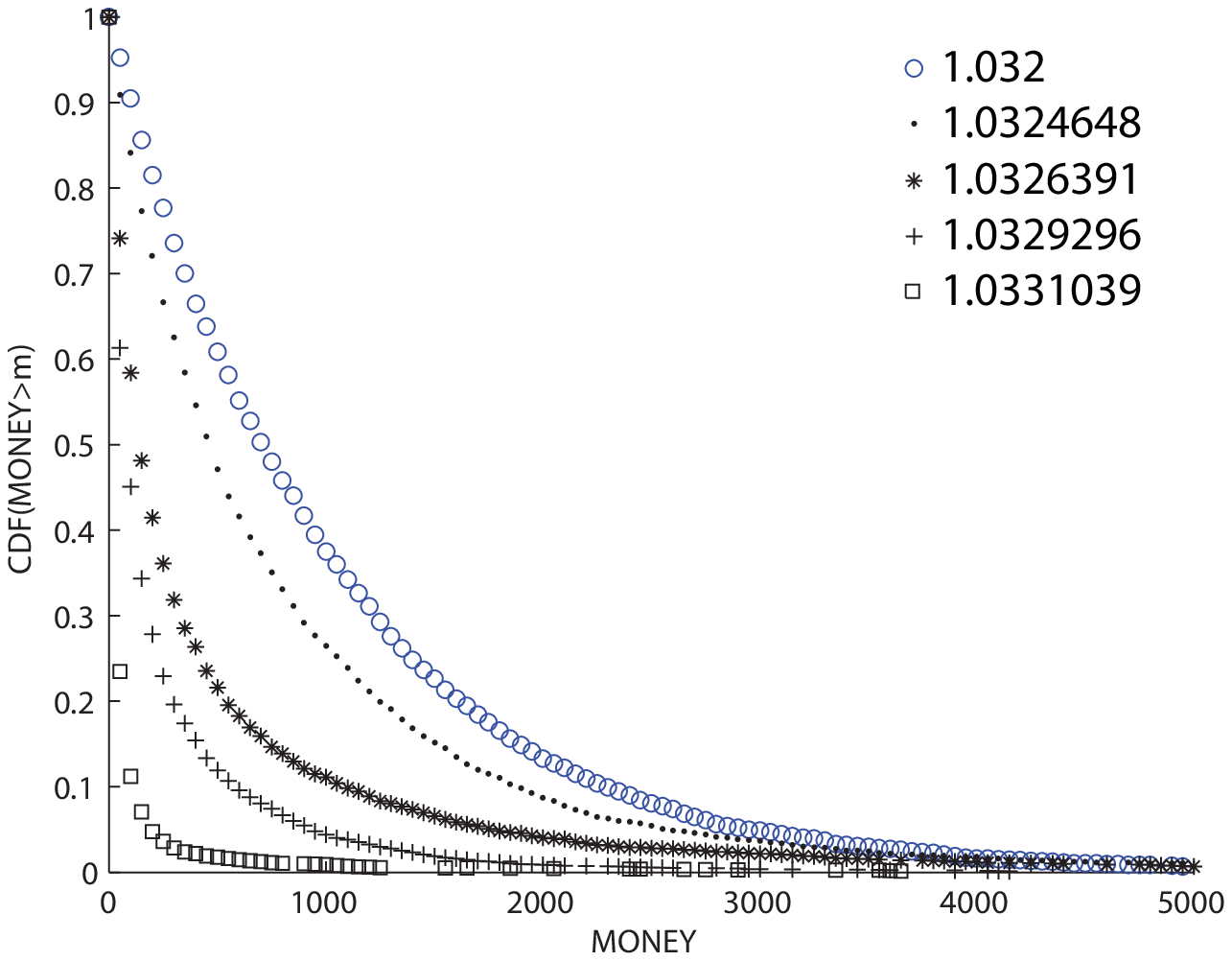} 
	\hskip 0mm \includegraphics[width=0.5\textwidth]{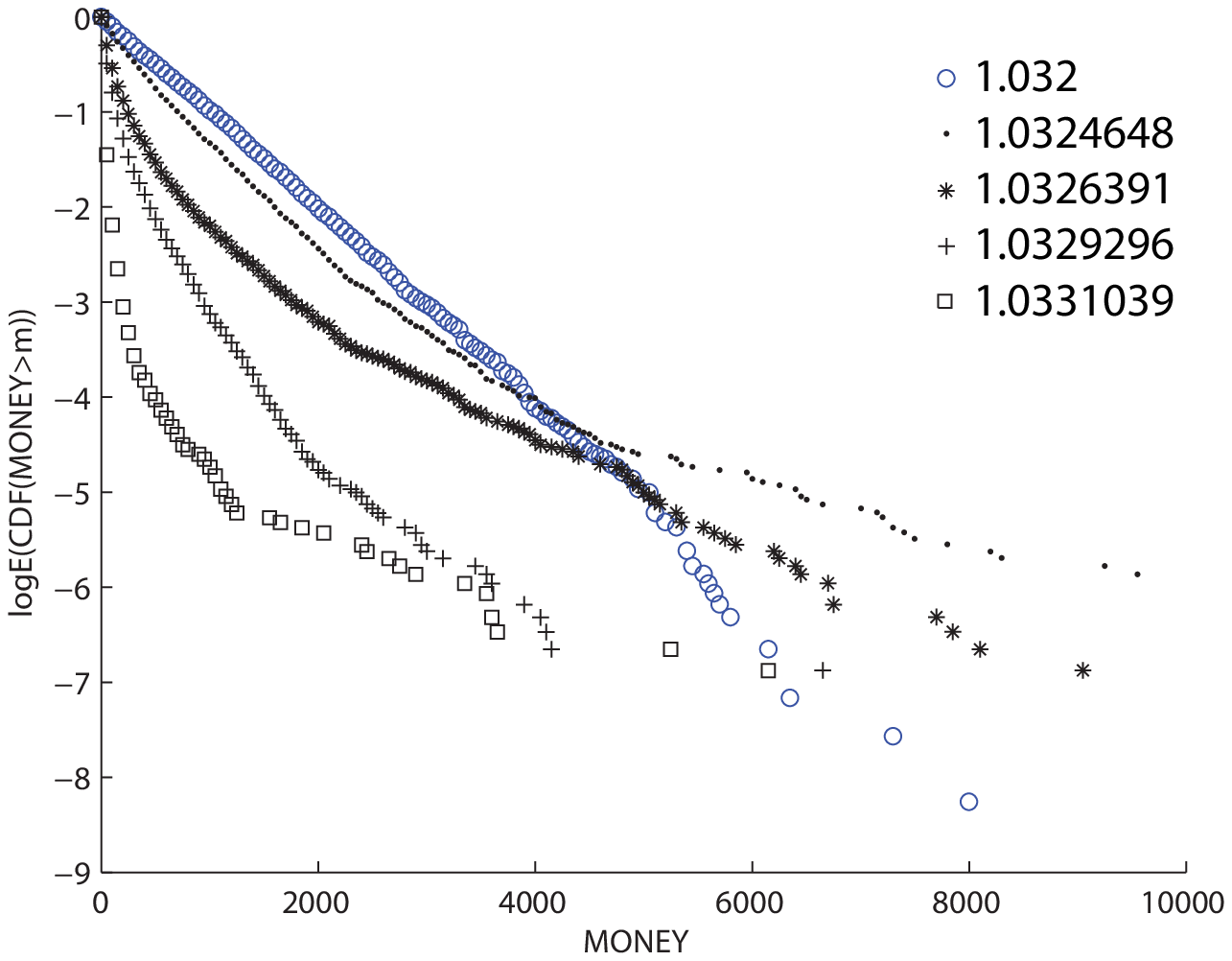}
	\hskip 0mm \includegraphics[width=0.5\textwidth]{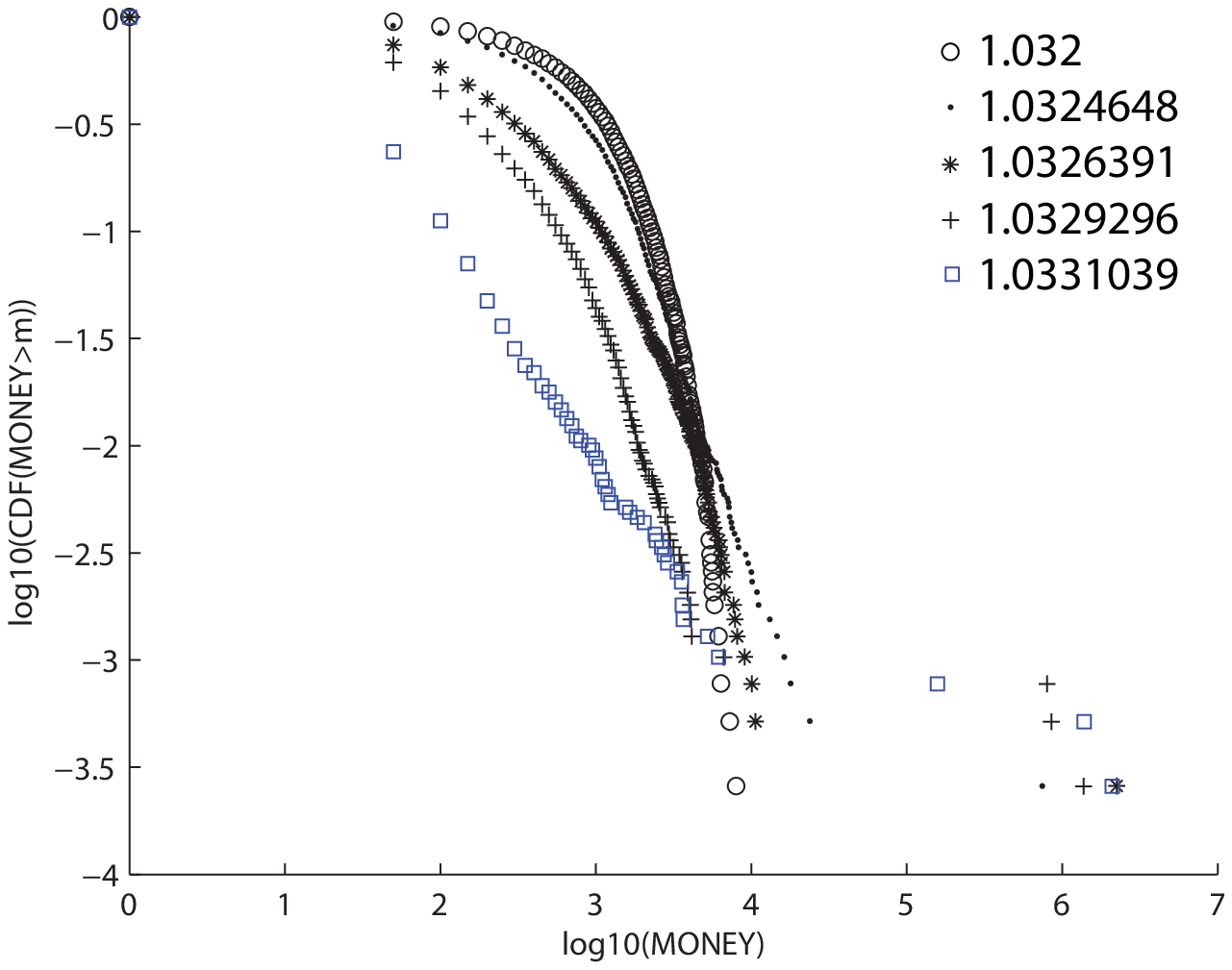}}
  \caption{Final CDF's obtained in the simulations for different axis scales. 
  (a) Final CDF's for parameters $\lambda_a = 1.032$ and $\lambda_b = 1.032, 1.324648, 1.326391, 
  1.329296$ and $1.331039$ in natural units. (b) Representation of the same CDF's in a logarithmic plot. 
  (c) Representation of the same CDF's in a double logarithmic plot.}
  \label{Fig2}
\end{figure}

After removing passive agents and their money, the final distributions of money are shown 
in Fig. \ref{Fig2} with different axes scales and for different values of $\lambda_b$. 
Here $\lambda_a = 1.032$ for all cases. As it can be seen, the symmetric case produces an
exponential distribution, as in the random scenario of the gas-like model [5]. Amazingly, 
increasing the asymmetry of the chaotic selection, the money distribution degenerates 
progressively from the exponential shape to a power law profile.

Fig.\ref{Fig2} (a) shows in five curves, the cumulative distribution functions (CDF's) obtained 
for the following four cases: $\lambda_a = 1.032$  and  $\lambda_b = 1.032, 1.324648, 1.326391, 
1.329296$  and  $1.331039$. Here the probability of having a quantity of money bigger or equal to 
the variable MONEY, is depicted in natural axis plot. The symmetric case $\lambda_a = \lambda_b =1.032$ 
is highlighted in blue color. Fig.\ref{Fig2} (b) shows the same data in a natural logarithmic plot, 
to illustrate the exponential profile obtained in the symmetric case (highlighted in blue). It also 
shows how a progressive asymmetry in the selection of agents degrades this characteristic profile. 
Fig.\ref{Fig2} (c) shows again the same data but in a double-logarithmic plot. It gives an extensive 
view of the degradation observed before, making evident the straight line power-law profile obtained 
in the most asymmetric case (highlighted in blue).

As a consequence, one may say that a chaotic selection of agents is reproducing the two characteristic 
distributions observed in real wealth distributions. On one hand, one obtains the exponential distribution, 
as in the random gas-like models. This characteristic appears when chaotic market is operating under 
a symmetric rule of selection of interacting agents. The significance may be that the correlations in 
agent interactions looks like random, and all agents win or lose with fair probability.

On the contrary, when asymmetry is introduced in the chaotic process of selection of agents, the 
distribution of money becomes progressively more unequal. The probability of finding an agent in 
the state of poorness increases and only a minority of agents reach very high fortunes. What 
is happening here is, that the asymmetry of the chaotic map is selecting a set of agents preferably 
as winners for each transaction ($j$ agents). While others, with less chaotic luck, become preferably 
the losers ($i$ agents). As in real life, there are markets where some individuals possess a preferred 
status and this makes them win in the majority of their transactions.

From these primary results, it seems interesting to study the dynamics of the system at micro detail. 
This might help to uncover the possible causes that originate unfair distributions of money in society. 
Then, simulations are repeated for a more tractable number of agents $N=500$. The initial money is 
$m_0=1000\$$  as previously, and the simulations take a total time of $T=2*N^2 = 0.5$ millions of transactions.

\begin{figure}
  \centerline{
    \includegraphics[width=0.5\textwidth]{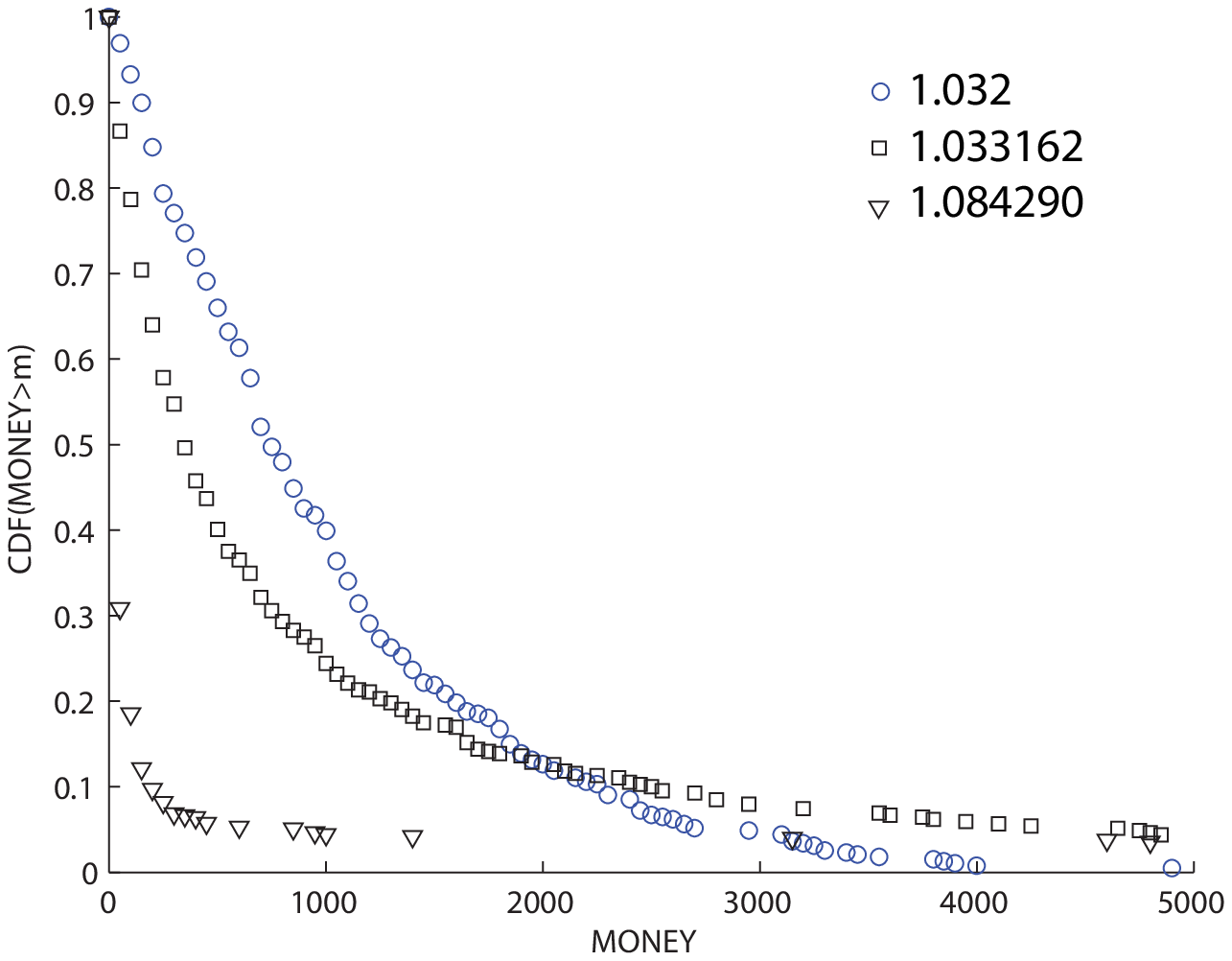} 
	\hskip 0mm \includegraphics[width=0.5\textwidth]{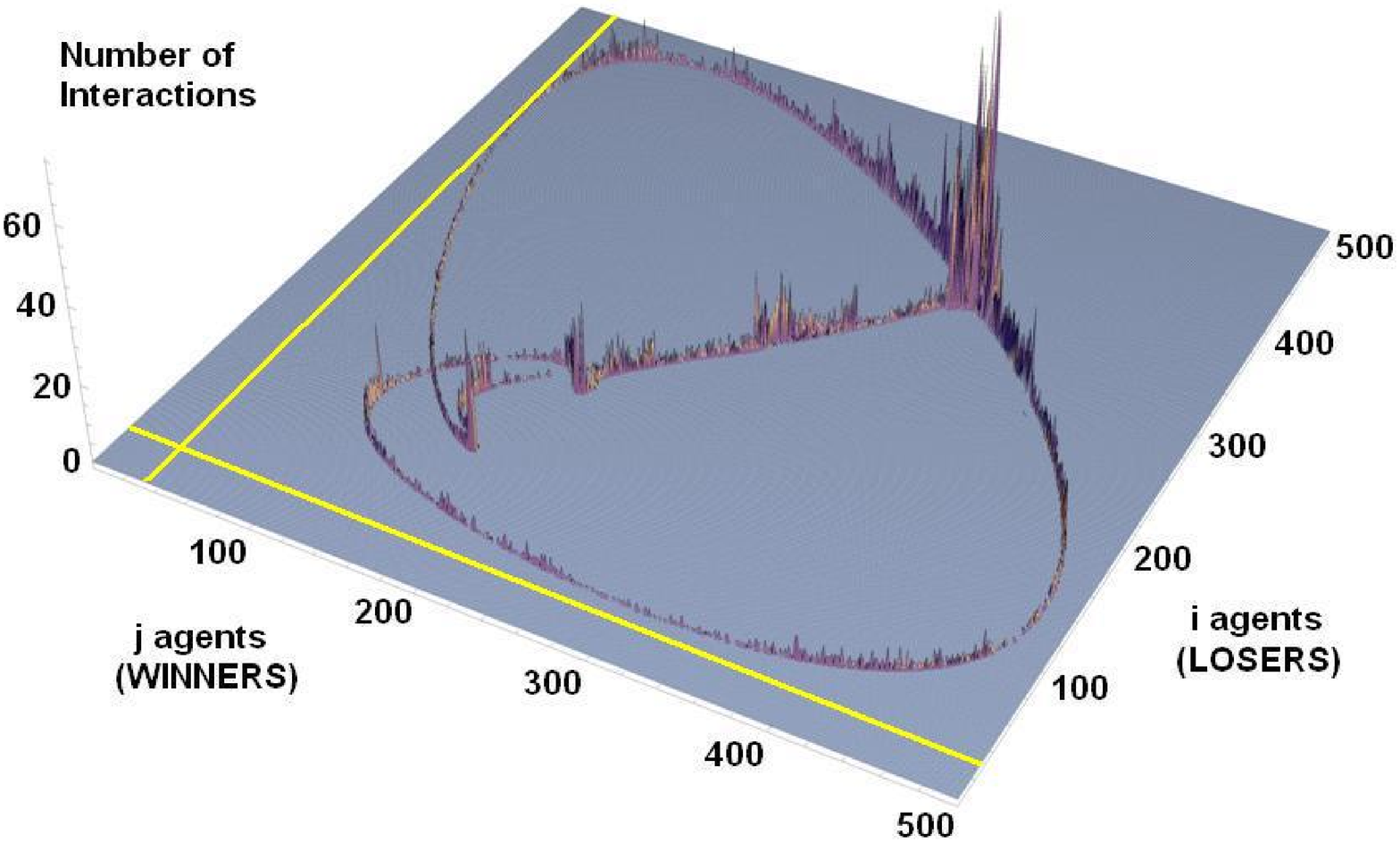} 
	\hskip 0mm \includegraphics[width=0.5\textwidth]{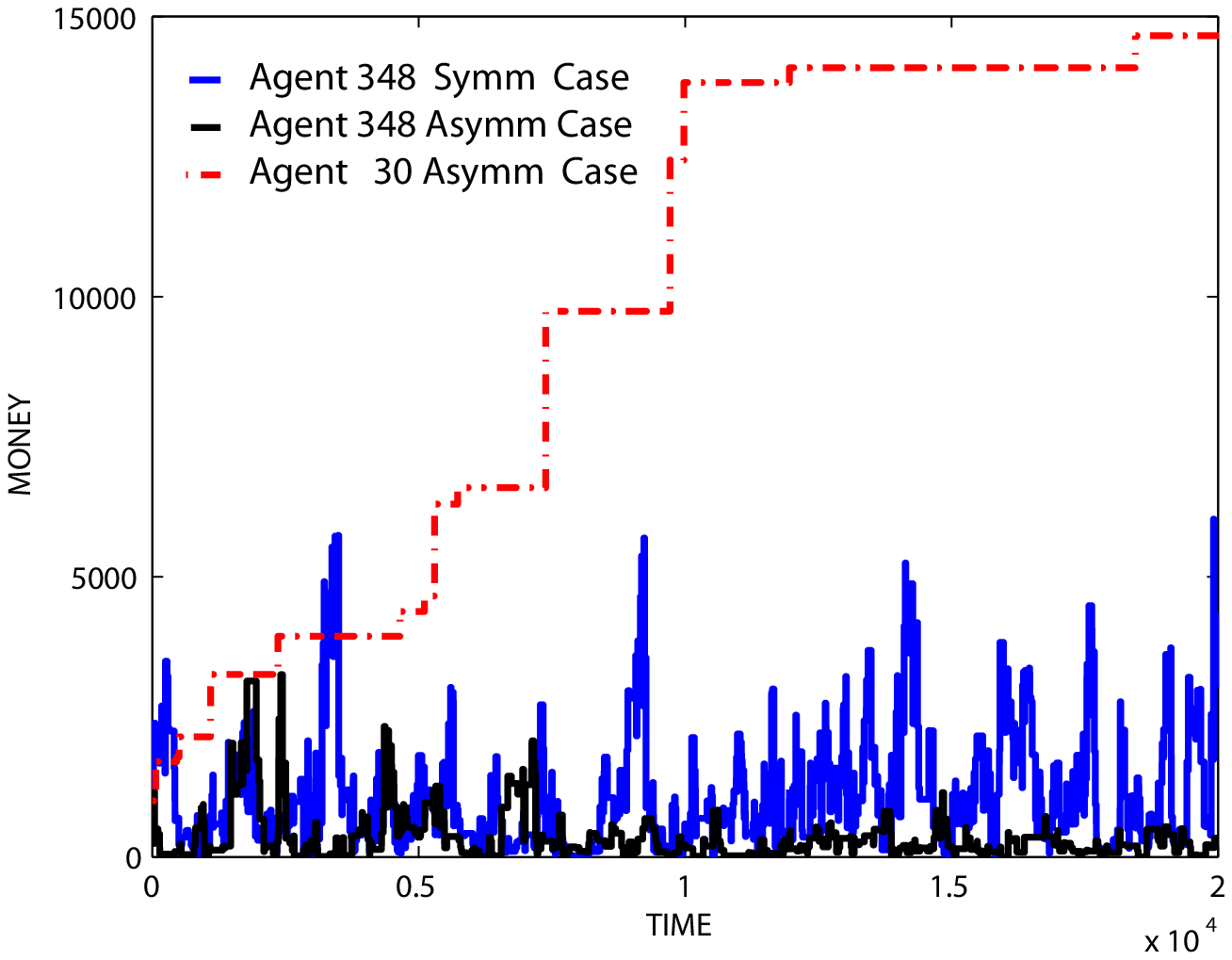}}
  \caption{Results obtained for a simulation scenario with $N=500$ agents. (a) Representation 
  of the same CDF's of Fig.\ref{Fig2} but for $N=500$ agents and three different cases 
  ($\lambda_b = 1.032, 1.033162$ and $1.08429$). (b) Number of transactions between pairs of agents 
  for the most asymmetric case ($\lambda_a = 1.032$ and $\lambda_b = 1.08429$). Axis $x$ and $y$ 
  represent pairs of agents $(j,i)$, selected as winner ($j$) or loser ($i$) respectively. Axis $z$ 
  represents the number of transactions of each pair. Agent $30$ position is highlighted with a yellow 
  line to show that he never looses ($i$ agent). (c) Time evolution of money for agent $348$ in the 
  symmetric case ($\lambda_a = \lambda_b = 1.032$), and agents $348$ and $30$ in the most asymmetric 
  case ($\lambda_b = 1.08429$).}
  \label{Fig3}
\end{figure}

First the CDF is obtained for three simulation cases with different $\lambda_b$ values 
(see Fig.\ref{Fig3} (a)). The degradation of the exponential distribution is again obtained as 
asymmetry increases. However when Fig.\ref{Fig2} (a) and Fig.\ref{Fig3} (a) are compared an 
avalanche effect becomes now evident for bigger markets. Here one may appreciate that the shape 
of the money distributions in the symmetric case is the same for a simulation scenario of 
$N=5000$ or $N=500$ agents. In contrast, the asymmetric cases produce more unequal or dramatic 
distributions when the community of agents is greater. A slight change in the asymmetry of the 
interactions, compare for example the case $\lambda_b = 1.033162$, produces a significantly higher 
degradation in the final distribution of money for the $N=5000$ market. At this point, it is 
interesting to remark that, in a chaotic scenario the size of the market matters. A small asymmetry
 in the rules of selection of agents will become far more aggressive in its final distribution of 
 wealth depending on the size of the market.

As a consequence, the globalization of markets that evolve under chaotic rules, may suffer avalanche 
effects that drive more unequal and dramatic distributions of wealth. This may resemble some 
situations observed in real economy, where a small unbalance in global markets, may produce large 
differences in the share of capital among individuals.

Secondly, the network of chaotic economic interactions is analyzed. Agents trade in a complex network 
of interactions illustrated in Fig.\ref{Fig3} (b). As we can see, agents in the range $320$ to $360$ 
have quite a higher number of interactions between them. Amazingly, these agents are not specially 
treated in the symmetric case, for in this scenario all agents have the same opportunities. On the 
other hand, the figure shows how in the most asymmetric case some agents are preferably selected as 
winners. The maximum asymmetry case depicted here treats some agents as winners in all transactions. 
See highlighted in yellow, agent $30$, who is never selected as an i agent, a looser. In this situation 
the group of the agents in the range $320$ to $360$, which interact mainly among themselves, will always 
get poorer as the majority of the society is getting poor. The more they interact the sooner they will 
lose all their money. Put it in another way, in the asymmetric case, not all agents have the same opportunities.

In third place, Fig.\ref{Fig3} (c) shows the evolution in time of an agent's money in symmetric and 
asymmetric cases. Here agent number $348$, who in the range $320$ to $360$, is depicted as an example. 
In the picture it can be observed that when the market is symmetric, agents' wealth oscillates in a 
random-like way. Becoming rich or poor in the end it is just a question of having a very profitable 
transaction, being in the right place at the right time. In the chaotic market model this is simply 
the result of some specific selection of the initial conditions and chaotic parameters. Let us note 
that in the symmetric case, the money distribution becomes exponential. On a global perspective, there 
are no sources or sinks of wealth and even then, when the individuals have the same opportunities, 
the final distribution is unequal.

However when the market is asymmetric agent $348$ becomes inevitably poorer as time passes. In this 
case there are agents that never lose (as for example, agent $30$) and they become inevitably sinks 
of wealth. This can be observed in Fig.\ref{Fig3} (c) where agent $30$ is precisely one of these sinks, 
accumulating more and more money every time he trades. These chaotic favored agents make the rest 
of the community poorer. On a global perspective, becoming rich in the asymmetric case depends on who 
you are and who you interact with (the shape of the chaotic attractor and its initial conditions). 
This resembles a society where some individuals belong to specific circles of economic power.

\begin{figure}
  \centerline{
    \includegraphics[width=0.5\textwidth]{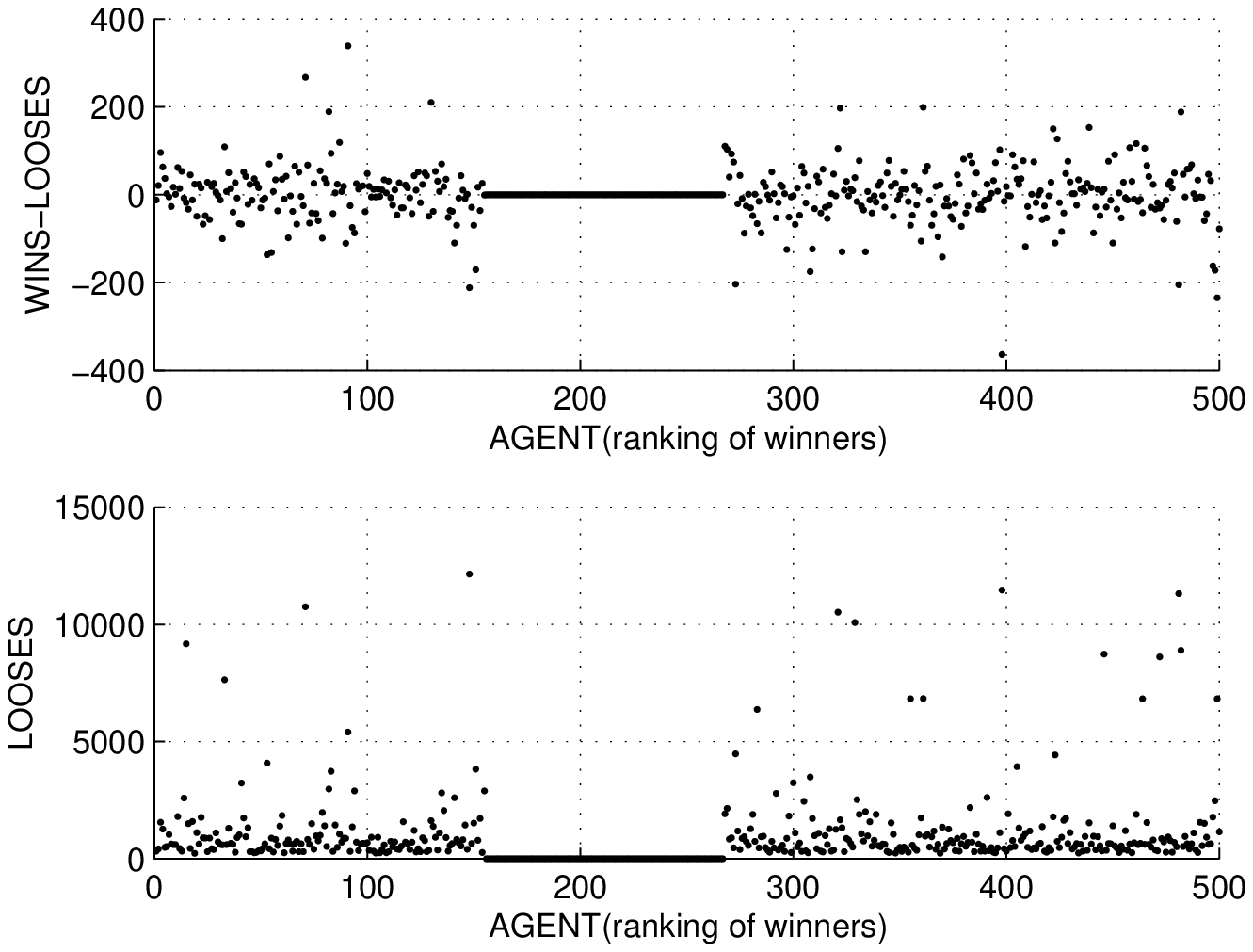} 
	\hskip 3mm \includegraphics[width=0.5\textwidth]{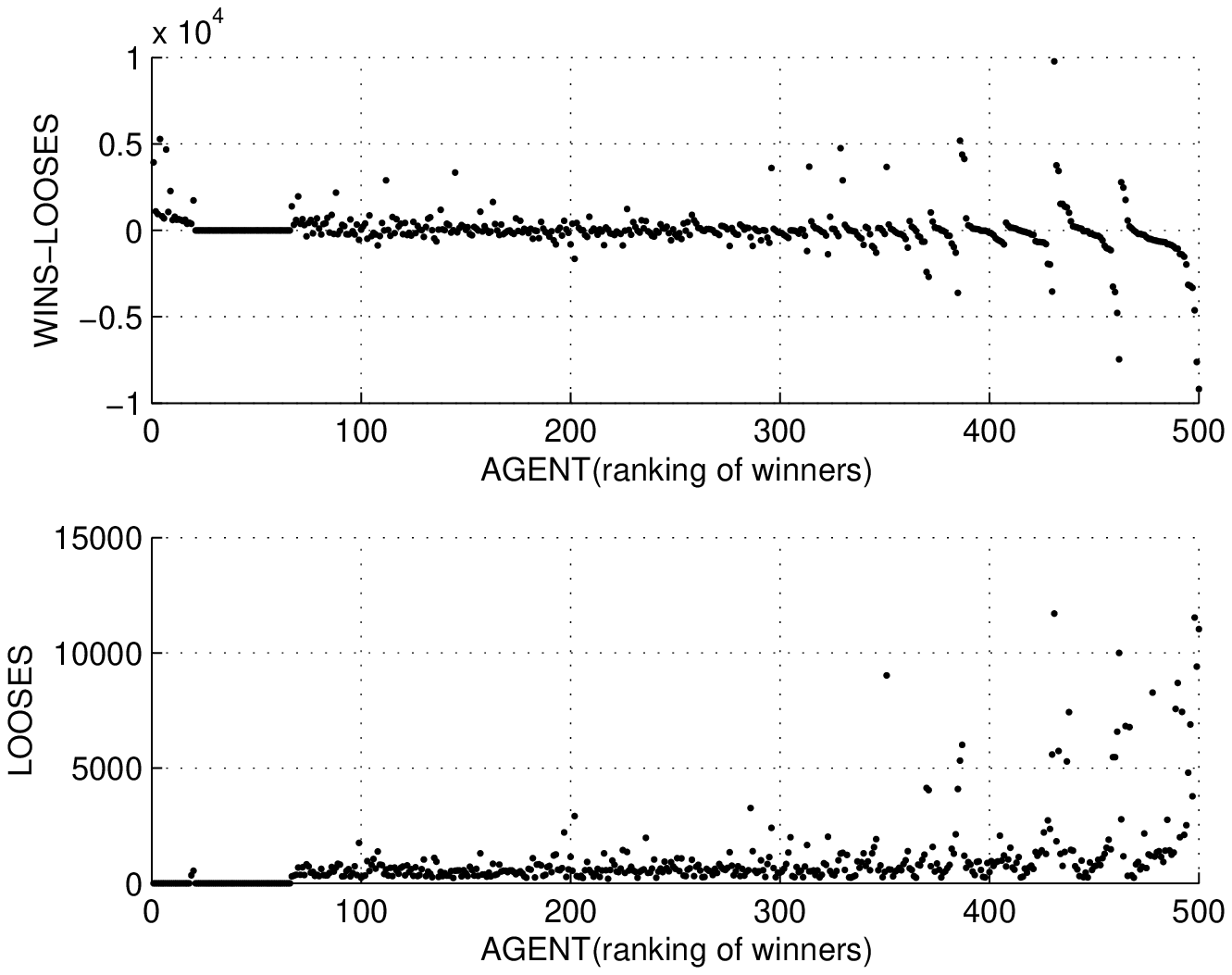}}
  \caption{Representation of the role of $N=500$ agents in the community after all interactions. 
  Agents are arranged in descending order according to their final wealth (agent $0$ is the richest 
  of all). The upper graphic shows the total number of wins over looses of an agent. The bottom 
  graphic shows the number of times an agent has been selected as i agent or looser. 
  (a) Symmetric case with $\lambda_a = \lambda_b = 1.032$. Here, the number of wins and looses 
  is uniformly distributed among the community. A range of $112$ individuals are passive and 
  always have the initial $1000\$$. (b) Asymmetric case with $\lambda_a = 1.032$  and 
  $\lambda_b = 1.084290$. Here, $46$ agents never interact and $18$ agents never loose.}
  \label{Fig4}
\end{figure}

Finally, to illustrate these qualitative observations, we study the number of times that agents 
win or lose in the simulation cases. The results obtained are depicted in Fig.\ref{Fig4}. Here it 
is shown the number of times (interactions) that an agent has been a looser (bottom graph) and the 
difference of winning over losing times (top graph). The $x$ axis shows the ranking of agents 
arranged in descending order according to their final wealth. So, agent number $0$ is the richest 
of the community and agent number $500$ is the poorest.

Fig.\ref{Fig4} (a) depicts the symmetric case, where $\lambda_a = \lambda_b =1.032$. Here, 
the number of wins and looses is uniformly distributed among the community. There is also a 
range of agents that don't interact ($112$ agents), this can be seen clearly in the figure now. 
In this case, the chaotic selection of agents shows no particular preference for any other and 
the final distribution becomes the exponential. This is similar to traditional gas-like simulations 
with random agents [~\refcite{4} -~\refcite{9}].

Fig.\ref{Fig4} (b) shows the same magnitudes for the asymmetric case with $\lambda_a = 1.032$  
and $\lambda_b = 1.0842908$. Here the asymmetry is maximum prevailing chaotic behavior. In this 
case, there is a group of agents in the range of maximum richness that never loose. The chaotic 
selection is giving them maximum luck and this makes them richer at every transaction. A range 
of agents, lower than in the symmetric case, are passive and never interact ($46$ agents). In 
this case, the final money distribution becomes quite unequal: $269$ agents (half of the society 
approximately) end half poorer with a final wealth inferior to $500\$$ and of them, and $56$ 
agents (approx. the $10\%$) finish with less that $50\$$. It is also interesting to see in 
Fig.\ref{Fig4} (a) that in the poor class, there are agents that have a positive difference of 
wins over looses, but amazingly they become poor anyway. Consequently, one can deduce that they 
are also bad luck individuals. They may be selected as $j$ agents in most part of their transactions, 
but unfortunately their corresponding trading partners ($i$ agents) are poor too, and they can 
effectively earn low or no money in these interactions.

To summarize, these results show that the symmetric case of selection of agents resembles a 
society where agents have equal opportunities in the market. In this case, agents are equally 
selected winners or losers (symmetry, $x \longleftrightarrow y$ and  $i \longleftrightarrow j$) 
in a chaotic and correlated fashion. This resembles a market where all agents have the same 
opportunities of success. In this case, there is no particular preference for any other and the 
final distribution of money becomes exponential, as in the random gas-like model.

Quite the opposite, when a small asymmetry is introduced in the rule of selecting agents, the 
society becomes more unequal. Some agents become preferred winners for most of their transactions 
and the money. The opportunities of winning in the end just depend on interacting with the proper 
group of rich agents. This resembles a market where there are established circles of power and 
corruption. Additionally, the chaotic nature of the market may also reproduce an avalanche effect 
as a result of globalization of markets, where a small asymmetry may produce extreme inequality.

\section{CONCLUSIONS}

This work introduces a novel approach in the field of economic (ideal) gas-like models 
[~\refcite{4} -~\refcite{9}]. It proposes a chaotic selection of agents in a trading community 
and it accomplishes to obtain the two most important money distributions observed in real 
economies, exponential and power-law distributions, besides a mechanism of transition between them.

The introduction of chaos is based on two considerations. One is that real economy shows a 
kind of a chaotic character. The other one is that dynamical systems offer a very simple and 
flexible tool able to break the inherent symmetry of random processes and overcome the technical 
difficulties of gas models to shift from the Boltzmann-Gibbs to the Pareto distributions.

A specific 2D dynamical system under chaotic regime is considered to produce the chaotic 
selection of agents. This system represents two logistic maps coupled in a multiplicative 
way and produces a strong correlation between the interacting agents. This correlation 
guaranties a complex behavior that can be tuned through two system parameter ( $\lambda_a$ and $\lambda_b$) 
and so, it is able to produce symmetric or asymmetric conditions in the market.

The results obtained in this model show how a chaotic selection of agents under symmetric conditions 
produces a final exponential distribution of money. In contrast, when asymmetry is introduced in the 
selection of agents, the distribution of money becomes progressively more unequal until it produces 
power law profiles.

The analysis of micro level interactions in both cases shows what is happening in the market and what 
is responsible of producing these two global phases (exponential or power law).

The main conclusions produced by this model may resemble some situations in real economy. 
In a symmetric scenario the selection of agents resembles a society where agents have equal 
opportunities in the market. In this case, there is no particular preference for any other and 
the final distribution becomes the exponential. This is similar to traditional gas-like 
simulations with random agents. On a global perspective, there are no sources or sinks of 
wealth and even when the individuals have the same opportunities, the final distribution is 
unequal. From an agent point of view, becoming rich or poor just depends of being at the right 
time in the right place.

In the asymmetric scenarios, a small asymmetry is introduced in the rule of selecting agents 
and then, the society becomes unequal. Some agents become preferred winners for most of their 
transactions. The opportunities of winning in the end just depend on being a preferred agent 
and interacting with the proper group of rich agents. The final money distribution resembles 
the Pareto distribution. Moreover, the chaotic nature of the market also reproduces the impact 
of the size of the market on the final distribution of wealth. When unequal or asymmetric 
conditions of trading are ruling, an avalanche effect may be produced and then inequality 
becomes extreme.

The authors hope that this work that may bring new ideas and perspectives to Economy and 
Econophysics. The proposal of considering chaotic dynamics in multi-agents modelling may also 
be of interest to other fields, where scientists try to describe and understand complex systems.

\section*{Acknowledgments}
The authors acknowledge some financial support by Spanish grant DGICYT-FIS2009-13364-C02-01.




\begin{thebibliography}{00}
\bibitem{1} T. Lux, F. Westerhoff, Economics crisis, {\it Nature Physics}, {\bf 5}, 2 January (2009).
\bibitem{2} J. Doyne Farmer, D. Foley, The economy needs agent-based modelling, {\it Nature}, {\bf 460}, 
685-686 August (2009).
\bibitem{3} R. Mantegna, H.E. Stanley, {\it An Introduction to Econophysics: Correlations 
and Complexity in Finance},ed. Cambridge University Press, 2000, ISBN 0521620082.
\bibitem{4} A. Chatterjee, S. Yarlagadda, B. Chakrabarti, {\it Econophysics of Wealth Distributions}, 
ed. Springer, 2005.
\bibitem{5} V.M. Yakovenko, {\it Statistical Mechanics Approach to}, Encyclopedia of Complexity 
and System Science, Econophysics, ed. Springer, 2009, ISBN 9780387758886.
\bibitem{6} A. Dragulescu, V.M. Yakovenko, Statistical Mechanics of Money, {\it The European 
Physical Journal B}, {\bf 17}, 723-729 (2000).
\bibitem{7} J.P. Bouchaud, M. Mezard, Wealth Condensation in a Simple Model Economy, {\it Physica A}, 
{\bf 282}, 536-545 (2000).
\bibitem{8} B.K. Chakrabarti, S. Marjit, Self-organization in Game of Life and Economics, 
{\it Indian Journal Physics B}, {\bf 69}, 681-698 (1995).
\bibitem{9} A. Chakraborti, B.K. Chakrabarti, Statistical mechanics of money: how saving 
propensity affects its distribution, {\it The European Physical Journal B}, {\bf 17}, 167-170 (2000).
\bibitem{10} A. Christian Silva, V.M. Yakovenko, Temporal evolution of the "thermal" and 
"superthermal" income classes in the USA during 1983-2001, {\it Europhys. Lett.}, {\bf 69 (2)}, 304-310 (2005).
\bibitem{11} A. Dragulescu, V. M. Yakovenko, Evidence for the exponential distribution 
of income in the USA, {\it The European Physical Journal B}, {\bf 20}, 585 (2001).
\bibitem{12} A. Dragulescu, V.M. Yakovenko, Exponential and Power-law Probability Distributions 
of Wealth as Income in the United Kingdom and the United States, {\it Physica A}, {\bf 299}, 213 (2001).
\bibitem{13} V. Pareto, {\it Cours d'Economie Politique}, ed. F. Rouge (Lausanne and Paris, 1897)
\bibitem{14} S. Sinha, Evidence of the Power-law Tail of the Wealth Distribution in India, 
{\it Physica A}, {\bf 359}, 555-562 (2006).
\bibitem{15} A.Y. Abul-Magd, Wealth Distribution in an ancient Egyptian Society, {\it Phys. Rev. E}, 
{\bf 66}, 057104 (2002).
\bibitem{16} BK. Chakrabarti, A. Chatterjee, Ideal Gas-Like Distributions in Economy: Effects of 
Saving Propensity, in {\it Applications of Econophysics}, Proc 2nd Nikkey Econophys. Symp. 
Ed TakayasuH, Springer, Tokio, 280-285 (2004).
\bibitem{17} A. Corcos, J.-P. Eckmann, A. Malaspinas, Y. Malevergne and D.  Sornette, Imitation 
and contrarian behavior: hyperbolic bubbles, crashes and chaos, {\it Quantitative Finance}, 
{\bf 2}, 264-281 (2002).
\bibitem{18} J. Gonzalez-Estevez, M. Cosenza, O. Alvarez-Llamoza, R.Lopez-Ruiz, Transition from 
Pareto to Boltzmann-Gibbs Behavior in a Deterministic Economic Model, {\it Physica A}, {\bf 388}, 
3521-3526 (2009).
\bibitem{19} C. Pellicer-Lostao, R. Lopez-Ruiz, A chaotic gas-like model for trading markets, 
{\it Journal of Computational Science}, {\bf 1}, 24–32 (2010).
\bibitem{20} R. Lopez-Ruiz, R. Perez-Garcia, Dynamics of Maps with a Global Multiplicative Coupling, 
{\it Chaos, Solitons and Fractals}, {\bf 1}, 511-528 (1991).
\bibitem{21} C. Pellicer-Lostao, R. Lopez-Ruiz, Orbit Diagram of the H\`{e}non Map and Orbit 
Diagram of Two Coupled Logistic Maps, {\it The Wolfram Demonstrations Project}, 
http://demonstrations.wolfram.com
\end{thebibliography}
\end{document}